\documentclass[global,twocolumn]{svjour}
\usepackage{graphics}
\usepackage{amsmath}
\journalname{Applied Physics B}
\begin{document}
\title{Fast polarimetry system for the application to spin quantum non-demolition measurement}
\author{M. Takeuchi\inst{1}\and T. Takano\inst{1}\and
S. Ichihara\inst{1}\and A. Yamaguchi\inst{1}\and M. Kumakura\inst{1}\inst{2}\inst{3} \and T. Yabuzaki\inst{4} \and and Y. Takahashi\inst{1}\inst{3}
\thanks{Fax:+81-75-753-3769, E-mail:yitk@scphys.kyoto-u.ac.jp}
}                     
%
%
\institute{
 Department of Physics, Graduate School of Science, Kyoto University, Kyoto 606-8502, Japan
 \and
 PREST, JST, 4-1-8 Honcho Kawaguchi, Saitama 332-0012, Japan
 \and
 CREST, JST, 4-1-8 Honcho Kawaguchi, Saitama 332-0012, Japan
 \and
 Faculty of Information Science and Arts, Osaka Electro-Communication University, Osaka 572-8530, Japan
}
\date{Received: / Revised version:}
%
\maketitle
\begin{abstract}
We report the development of a fast pulse polarimeter
 for the application to quantum non-demolition measurement
 of atomic spin (Spin QND).
The developed system was
 tunable to the atomic resonance of ytterbium atom and
 has narrow laser linewidth
 suitable for spin QND.
Using the developed polarimeter,
 we successfully demonstrated the measurement of the vacuum noise,
 with $10^6$ to $10^7$ photon number per pulse.

\end{abstract}
\section{Introduction}

Recently, interest in the quantum nature of atomic ensembles
 has been growing. 
To measure the quantum nature of the atomic ensembles, 
 one of the most useful interaction is paramagnetic Faraday
 rotation of light \cite{happer72}.
Under some conditions,
 the probing of the atomic ensembles via the paramagnetic Faraday rotation
 with a polarimeter becomes 
 quantum non-demolition measurement of spin (spin QND)
 \cite{kuzmich98,takahashi99,kuzmich99}.

Spin QND is not only a quantum measurement,
 but also has a wide variety of the applications.
For example, spin squeezing, entanglement of two macroscopic objects, 
 and quantum memory for light have been demonstrated
 \cite{kuzmich00,julsgaard01,geremia04,julsgaard04}.
These application of spin QND will be some breakthroughes
 not only in precision measurements but also quantum information
 processings.

If we can perform the spin QND successively on the same atomic ensemble system,
 various interesting applications
 such as quantum teleportation would become possible.
For such purposes,
 each probe time $T$ must be much shorter than
 the spin coherence time of the system $T_2$.
Moreover,
 the probe laser should be tunable to a particular atomic resonance, 
 and the detuning $\Delta$ of the probe laser frequency from atomic resonance
 should be larger than the atomic linewidth $\Gamma$;
 $\Delta>\Gamma=\tau_\mathrm{e}^{-1}$ is required,
 where $\tau_\mathrm{e}$ is the lifetime of the excited state,
 because the decoherence due to the photon scattering should be avoided
 \cite{duan00}.
The bandwidth of the probe laser $\Gamma_\mathrm{p}$ needs not to be
 narrower than the linewidth $\Gamma$,
 but in the case of $\Gamma_\mathrm{p}<\Gamma$,
 we can safely ignore the effect of finite bandwidth of the laser
 in the analysis at any detunings.
Therefore, we adopt the conditions $\tau_\mathrm{e}<T\ll T_2$
 and $\Gamma_\mathrm{p}<\Gamma<\Delta$
 for ideal spin QND.
To the best of our knowledge, the most shortest probe time $T$
 in the applications of the spin QND is $T=\mathrm{50\mu s}$
 in Ref. \cite{geremia04},
 where $\tau_\mathrm{e}=\mathrm{30ns}$.
Although the polarimeter for the pulse of duration
 $T$ of much less than one nano-second
 has developed by use of pulse lasers
 for the interest in the quantum cryptography
 \cite{hansen01,hirano03},
 it is, however, inappropriate for spin QND,
 because the pulse laser is not tunable to a certain atomic
 resonance,
 and has wide frequency bandwidth compared with the natural full linewidth
 $\Gamma$.
Therefore, it is necessary to newly develop suitable
 detector as well as the light source.

In this paper, we report the development of tunable and narrowband fast
 polarimetry system
 suitable for the application to spin QND,
 and also report the measurement of the vacuum noise
 to demonstrate the performance.
The pulse width $T$ could be as short as 100 ns,
  nearly three order of magnitude improvement
  from the previous experiment of Ref. \cite{geremia04}.
Especially, we designed our system tunable for ytterbium (Yb)
 atomic resonance of $\mathrm{{}^1S_0\to{}^1P_1}$ transition
 with the resonant wavelength of $399.8~\mathrm{nm}$,
 because Yb is one of the best sample for spin QND
 with long coherence time compared with alkali-atoms.
It should be noted that
 this system is qualitatively different from the
 previous ones 
 in which a frequency side-band was measured by using spectrum analyzers
 as in ref. \cite{kuzmich00}. 
Technically, the pulse-mode operation in which a single quantum measurement
 is performed for each pulse \cite{hansen01,hirano03}
 is usually difficult because the noise of the system
 must be quantum-noise limited over broad frequency bandwidths.

\section{Review of Spin QND}

\begin{figure}[h]
 \includegraphics{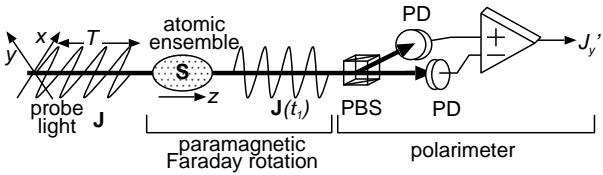}
 \caption{
  System of spin QND.
  The total spin component of an atomic ensemble along $z$ axis
   $S_z$ can be quantum non-demolitionally measured
   via the paramagnetic Faraday rotation by a polarimeter.
  The polarimeter consists of a polarization beam splitter (PBS),
  a photodiode (PD) and a differential amplifier.
 }
 \label{qnd}
\end{figure}

Firstly, let us review spin QND and the vacuum noise.
The interaction Hamiltonian of the paramagnetic Faraday rotation
 takes the form \cite{takahashi99},
\begin{equation}
    H=\alpha J_zS_z, \label{H}
\end{equation}
where $\alpha$ is a real constant.
$\mathbf{S}$ is the summation over the individual spin,
 which obeys the usual commutation relation of angular momenta
 $[S_i,S_j]=i\varepsilon_{ijk}S_k$,
 and the $z$ axis is set parallel to the wave vector of the probe
 light.
$\mathbf{J}$ is the quantum-mechanical Stokes vector of the light,
 which also obeys the usual commutation relation of angular momenta
 $[J_i,J_j]=i\varepsilon_{ijk}J_k$.
For a light pulse with the duration $T$ propagating in free space,
 $\mathbf{J}$ can be written as
 $J_x=\frac{1}{2}\int_0^T(a_+^\dagger a_-+a_-^\dagger a_+)dt$,
 $J_y=\frac{1}{2i}\int_0^T(a_+^\dagger a_--a_-^\dagger a_+)dt$,
 and
 $J_z=\frac{1}{2}\int_0^T(a_+^\dagger a_+-a_-^\dagger a_-)dt$,
 where $a_\pm$ is the annihilation operators of $\sigma_\pm$
 circular polarization mode, respectively \cite{duan00}.
From the backaction evasion $[S_z,H]=0$,
 this can be regarded as spin QND.
After the interaction time $t_1$,
 the $y$ component of the Stokes vector becomes
\begin{equation}
    J_y(t_1)\equiv e^{it_1H}J_ye^{-it_1H}\simeq J_y+\alpha t_1 J_xS_z,
    \label{J_y(t_1)}
\end{equation}
for $\alpha t_1 S_z \ll1$.
Therefore, by measureing $J_y(t_1)$,
 we can obtain information about $S_z$.
In Fig. \ref{qnd}, we show the schematic of spin QND.
We consider the case that the probe light is a coherent pulse
 which is linearly polarized along $x$ axis and
 contains $2J\gg 1$ photons on the average,
 and the polarimeter is set so as to measure the $y$-component
 of the Stokes vector $J_y(t_1)$.

The probability of obtaining the value $J_y'$ for the measurement of
 $J_y(t_1)$ is 
\begin{align}
 P(J_y') &\equiv \mathrm{Tr}_S\left(\langle J_y'|e^{-it_1H}\rho_{SJ}
 e^{it_1H}|J_y'\rangle\right),
 \label{P(J_y')}
\end{align}
 where $\mathrm{Tr}_S$ is the partial trace for the spin,
 $|J_y'\rangle$ is the eigenstate of $J_y(t_1)$ 
 with the eigenvalue of $J_y'$
 and $\rho_{SJ}$ is the density operator for the initial state
 of the spin and the probe light.
Without the atomic ensemble,
 $P(J_y')$ becomes \cite{duan00}
\begin{equation}
 P(J_y')\simeq\frac{\exp\left(-J_y'^2/(2\sigma^2)\right)}
 {\sqrt{2\pi\sigma^2}} \label{vacuum},
\end{equation}
 where $\sigma=\sqrt{J/2}$ is the standard deviation.
In this paper, we call this distribution of Eq.(\ref{vacuum})
 ``vacuum noise''.
With the atomic ensemble,
 the distribution $P(J_y')$ is broadened as
 $\sigma^2=J/2+(\alpha t_1J)^2\langle\Delta S_z^2\rangle$,
 which is easily derived from Eq. (\ref{J_y(t_1)}).
In general,
 the second term can become enlarge compared with the first term
 by prepareing a large number of atoms.
Therefore, if the vacuum noise can be measured
 with the polarimeter without atomic ensemble,
 which would be confirmed by the magnitude of $\sigma$,
 and its dependence on $J$,
 the performance of the polarimeter is usually sufficient
 in the application of spin QND \cite{geremia04}.
 
\section{Experimental Setup}

\begin{figure}[h]
 \begin{center}
  \includegraphics{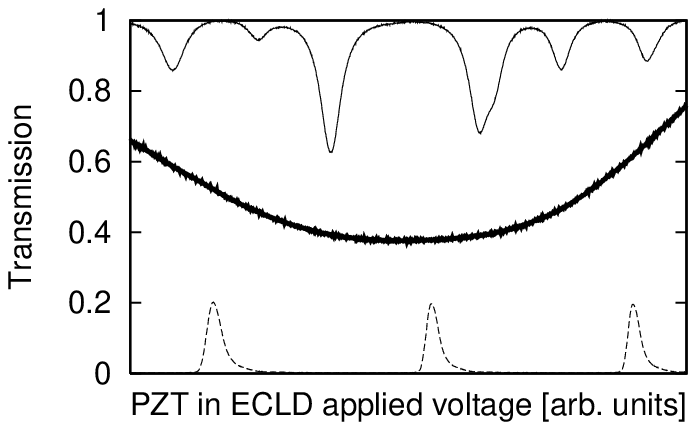}
  \caption{
    Transmission spectra of Yb atoms.
    The solid line is the transmittion of
     the atomic beam in the vacuum chamber.
    The bold line is the transmittion of the optogalvano cell,
     which is broadened due to the Doppler and pressure broadening
     effect.
    The dot line is the transmittion of the reference
     cavity with FSR=750 MHz.
  }
  \label{ecld}
 \end{center}
\end{figure}

As the CW light source, we constructed an external cavity laser diode
 (ECLD).
It provides a nominal CW output the power of about 10mW,
 where we use a GaN laser diode (Nichia NLHV3000E).
It is tunable around the resonant wavelength 399.8nm,
 and has the linewidth of several MHz, which is narrower than
 the full natural linewidth
 $\Gamma=2\pi\times29~\mathrm{MHz}$ \cite{komori03}.
To confirm that the laser frequency $\omega$ is
 fixed at the off-resonance of $\mathrm{{}^{171}Yb}$,
 we concurrently monitored the transmission of an optogalvano cell
 (Hamamatsu L2783-70AR-Yb),
 whose spectrum is well broadend by the Doppler effects and 
 the buffer gas collision.
In Fig. \ref{ecld},
 we show the transmission spectrum of the optogalvano cell,
 which is well broadened
 compared with the isotope shifts and hyperfine splittings
 \cite{banerjee03},
 as well as the transmission spectra of the Yb atomic beam
 and the reference cavity with FSR=750 MHz.

\begin{figure*}
 \begin{center}
  \includegraphics{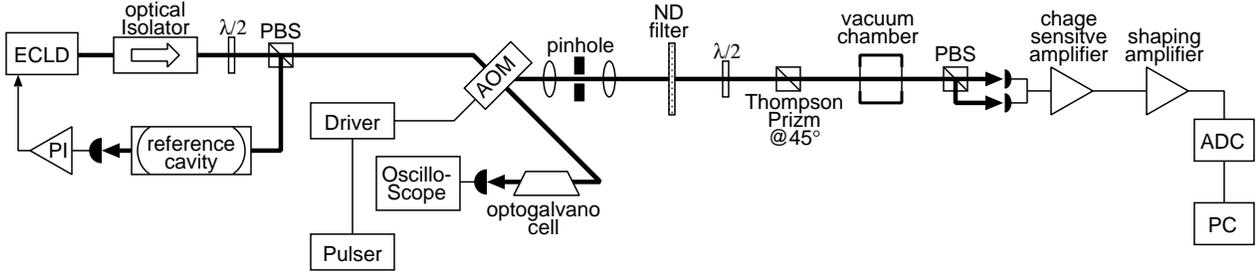}
  \caption{
    Experimental setup. As the CW source, we constructed an ECLD.
    The laser frequency $\omega$ was fixed at the off-resonance of
    $\mathrm{{}^{171}Yb}$ with a reference cavity and
     a PI amplifier \cite{riehle04},
     which was confirmed with the transmittion of an optogalvano cell.
    By use of an AOM,
     a pulse of the duration $T\sim100~\mathrm{ns}$ was generated.
    The pulse passed through an neutral density (ND) filter
     for the adjustment of the photon number per pulse $2J\sim 10^6$, 
     and a vacuum chamber, in which we can load atomic ensemble.
    After passing through the vacuum chamber, it was analyzed by
    a polarimeter.
  }
  \label{diagram.eps}
 \end{center}
\end{figure*}

To obtain an appropriate pulse from CW laser,
 we used a 1st-order diffracted light from an AOM (Isomet 1206C-833).
The pulse duration $T$ is variable by an pulser to the AOM.
To cut the higher order diffraction and to shape the transverse mode,
 the diffracted light from the AOM was passed 
 through two lenses and a pinhole.
To obtain ideal linear polarization along the $x$ axis,
 a $\lambda/2$ plate and a beamsplitting Thompson prism
 (Melles Griot 03PTB001) were also used,
 whose extinction ratio is $1\times 10^{-5}$.
As the adjustment of photon number per pulse $2J$,
 an neutral density (ND) filter was set in front of the prism.
We show these experimental setup in Fig.\ref{diagram.eps}.
We dare to perform the measurements using the vacuum chamber
 for the atom source to demonstrate robustness of our sysytem.
While the present paper describes the performance of the polarimetry system
 alone,
 however, not only the atomic beam
 but also the cold atoms are ready to be loaded in the vacuum chamber
 \cite{takeuchi05-2}.

\begin{figure}[h]
 \begin{center}
 \includegraphics{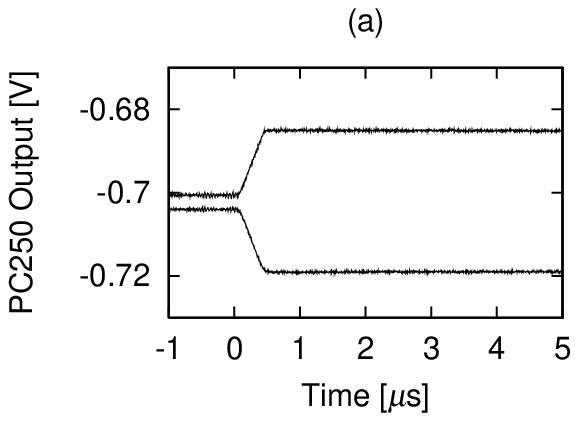}
 \includegraphics{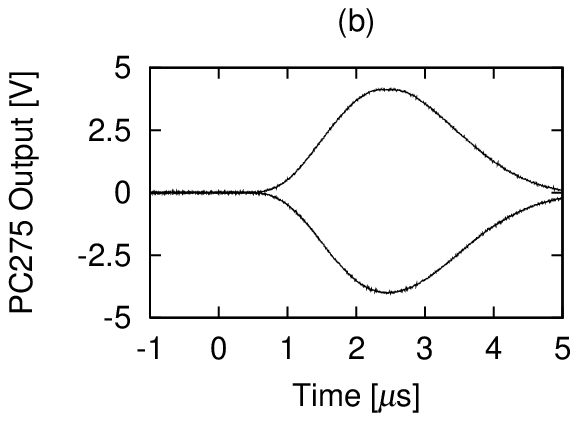}
 \caption{
 Responce for a pulse of the duration $T=400$ ns.
 To aquire these waveforms,
  we input a light pulse to the restictive photodiode.
 (a)Typical waveform of the charge sensitive amplifier.
  The sensitivity is $0.16~\mathrm{\mu V/electron}$.
 (b)Typical waveform of the shaping amplifier.
  The input waveform is the ones in Fig. \ref{amp} (a).
  The peak time is $2.3~\mathrm{\mu s}$ and the gain is
  typically $G=2\times 10^2$.
 }
 \label{amp}
 \end{center}
\end{figure}

As the photodetectors (PDs) in the polarimeter,
 we use sillicon PIN photodiodes (Hamamatsu S5973-02),
 with quantum effeciency $\eta=0.9$.
As the differential amplifier in the polarimeter,
 we use a charge sensitive amplifier (Amptek PC250) 
 and a shaping amplifier (Amptek PC275).
In Fig.\ref{amp}, we show the response for a light pulse
 of the duration $T=400$ ns.
The sensitivity of PC250 is
 $1/C_\mathrm{f}=0.16~\mathrm{\mu V/electron}$
 given by the feedback capacitance $C_\mathrm{f}=1~\mathrm{pF}$,
 and the tail length is $R_\mathrm{f}C_\mathrm{f}=300~\mathrm{\mu s}$
 given by the capacitance and
 the feedback resistance $R_\mathrm{f}=300~\mathrm{M\Omega}$.
The peak time of PC275 is $2.3~\mathrm{\mu s}$.
The gain of the shaping amplifier is typically set at
 $G=2\times 10^2$,
 as is shown in Fig. \ref{amp} (b).
The voltages at the peak time are read with an
 analog-digital converter (ADC, Innovative Integration AD40),
 which corresponds to $J_y'$ and
 the distribution of $J_y'$ are analyzed on the computer (PC).

\section{Measurement of the Vacuum Noise}

Here, we explain the procedure and results
 of the measurement of the vacuum noise.
Total photoelectron per pulse in the two PDs $2J$
 and the number difference between the two PDs $J_y'$,
 is obtained with the following procedure;
the average amplitude of the charge sensitive amplifier output
 corresponds to $J$ and was obtained with the light to one of the PDs blocked.
The proportional factor
 between the voltage and the photoelectron number is
 determined by the sensitivity of the PC250 $1/C_\mathrm{f}$.
The peak voltage of the shaping amplifier output
 corresponds to $J_y'$.
The proportional factor
 between the peak voltage and $J_y'$ is determined also by 
 the sensitivity of the PC250 and the gain of the shaping amplifier
 $G/C_\mathrm{f}$.

\begin{figure}[h]
 \begin{center}
 \includegraphics{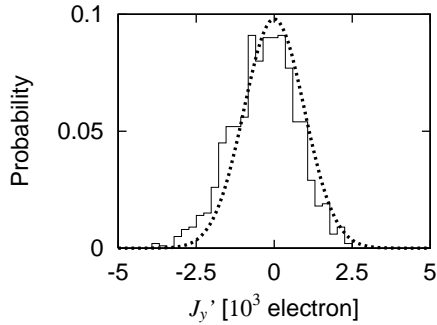}
 \caption{
  The distribution of the photon number difference $P(J_y')$
   at  (solid line).
  We constructed this distribution from the signal of 1000 pulses.
  Simultaneously we show the theoretical curve
  of Eq. (\ref{vacuum}) (dot line).
 }
 \label{hist}
 \end{center}
\end{figure}

\begin{figure}[h]
 \begin{center}
 \includegraphics{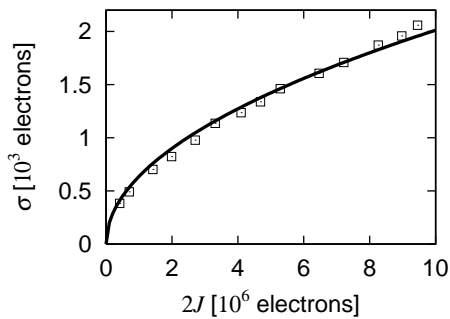}
 \caption{
  The standard deviation $\sigma$ of $P(J_y')$
   with various photon number per pulse (square).
  Thd solid line is the fitting curve $\sqrt{\epsilon J/2}$,
  where $\epsilon$ is the free parameter.
 }
 \label{variance}
 \end{center}
\end{figure}

In Fig. \ref{hist},
 we show the typical experimental result of $P(J_y')$,
 where we set $2J=3.7\times10^6$ and $T=400~\mathrm{ns}$.
This distribution well agree with the theoretical curve of
 Eq. (\ref{vacuum}).
To evaluate the photon number dependence of the distribution,
 we examined the standard deviation $\sigma$
 with various photon number per pulse $2J$.
In Fig. \ref{variance}, we show the values of $\sigma$ with
 various $2J$.
Simultaneously, we show the fitting curve $\sqrt{\epsilon J/2}$,
 where $\epsilon$ is the free parameter
 and set at $\epsilon=0.81$.
We think that the error is due to the precision of $C_\mathrm{f}$.
However, the dependence on $2J$ well agrees
 with $\sqrt{J/2}$.
The intrinstic excess noise was about $\sigma=8\times 10$,
 which is negligible compared with the vacuum noise
 for $10^6$ to $10^7$ photon number per pulse.
From these results,
 we concluded that
  we could successfully measure the vacuum noise.
It is also noted that we confirmed that the measurement
 of the vacuum noise is possible
 with the other photodiodes (Hamamatsu S7797, Hamamatsu S9687),
 and with other durations ($T=100$ ns , 200 ns, 600 ns).

Finally we note that toward the application to spin QND we have also
 succeeded in performing two-pulse polarimetry with pulse separation of
 as short as 5 $\mathrm{\mu s}$ by using two-independent polarimeters
 developed in this work.
The vacuum-noise limited performance was confirmed
 and no apparent classical correlation was observed between the readouts
 of the two polarimeters.

\section{Conclusion}

We successfully measured the vacuum noise
 of a light pulse from tunable CW laser for atom.
Especially, we constructed the system for ytterbium (Yb),
 because Yb is the best sample for spin QND.
The pulse duration $T\sim 100$ ns is much shorter
 than the previous experiments of the application of the spin QND
 \cite{julsgaard01,geremia04,julsgaard04},
 which will enhance the abilities of the quantum information processing
 with atomic ensemble.
Moreover, this technique is advantageous for reserach on
 quantum nature of atomic spin ensemble,
 such as squeezed spin state \cite{takeuchi05}.

\section{Acknoledgements}
We thank D. Komiyama for his experimental assistance.
We also thank important advices from T. Hirano.
This work was supported by the Strategic Information
 Communications R\&D Promotion Programme (SCOPE-S),
 and Grant-in-Aid for the 21st century COE,
  ''Center for Diversity and Universality in Physics''
  from the Ministry of Education, Culture, Sports, Science,
  and Technology (MEXT) of Japan.
M. Takeuchi is supported by JSPS.

\end{document}